\documentclass{ieeeaccess}
\usepackage{cite}
\usepackage{amsmath,amssymb,amsfonts,bm,mathtools}
\usepackage{amsthm}
\theoremstyle{plain}
\newtheorem{rmk}{Remark}
\usepackage{algorithmic}
\usepackage{graphicx}
\usepackage{enumitem}
\usepackage{url}
\usepackage{tikz}
\definecolor{accessblue}{cmyk}{1, 0.3, 0, 0.2}
\definecolor{greycolor}{cmyk}{0,0,0,.8}
\usetikzlibrary{arrows,shapes,shadows,fit,positioning}
\tikzstyle{int} = [thick, draw, fill=gray!10, drop shadow, minimum size=2em]
\tikzstyle{branch} = [fill, shape=circle, minimum size=3pt, inner sep=0pt]
\tikzstyle{multp} = [thick, draw, shape=circle, minimum size=3pt, inner sep=0pt]
\tikzstyle{port} = [draw, shape=circle, minimum size=3pt, inner sep=0pt]

\tikzstyle{block} = [draw, fill=blue!20,minimum size=2em]

\usepackage{textcomp}
\def\BibTeX{{\rm B\kern-.05em{\sc i\kern-.025em b}\kern-.08em
    T\kern-.1667em\lower.7ex\hbox{E}\kern-.125emX}}

\newcommand{\diff}{\mathrm{d}}

\begin{document}
\history{Date of publication xxxx 00, 0000, date of current version xxxx 00, 0000.}
\doi{10.1109/ACCESS.2017.DOI}

\title{%
    Trajectory Tracking Control of\\
    The Second-order Chained Form System\\
    by Using State Transitions%
}
\author{\uppercase{Mayu Nakayama}\authorrefmark{1}, \uppercase{Masahide Ito}\authorrefmark{1}, \IEEEmembership{Member, IEEE}}.
\address[1]{School of Information Science and Technology, Aichi Prefectural University, Nagakute, Aichi, Japan}

\markboth
{Nakayama~\headeretal: Preparation of Papers for IEEE Access}
{Nakayama~\headeretal: Preparation of Papers for IEEE Access}

\corresp{Corresponding author: Masahide Ito (e-mail: masa-ito@ist.aichi-pu.ac.jp).}
.

\begin{abstract}
    This paper proposes a novel control approach composed of sinusoidal reference trajectories and trajectory tracking controller for the second-order chained form system.
    The system is well-known as a canonical form for a class of second-order nonholonomic systems obtained by appropriate transformation of the generalized coordinates and control inputs. The system is decomposed into three subsystems, two of them are the so-called double integrators and the other subsystem is a nonlinear system depending on one of the double integrators. The double integrators are linearly controllable, which enables to transit the value of the position state in order to modify the nature of the nonlinear system that depends on them. Transiting the value to “one” corresponds to modifying the nonlinear subsystem into the double integrator; transiting the value to “zero” corresponds to modifying the nonlinear subsystem into an uncontrollable linear autonomous system. Focusing on this nature, this paper proposes a feedforward control strategy. Furthermore, from the perspective of practical usefulness, the control strategy is extended into trajectory tracking control by using proportional-derivative feedback. The effectiveness of the proposed method is demonstrated through several numerical experiments including an application to an underactuated manipulator.
\end{abstract}


\titlepgskip=-15pt

\maketitle

\section{Introduction}
\label{sec:introduction}

\PARstart{N}{onholonomic} systems are nonlinear dynamical systems with non-integrable differential constraints, whose control problems have been attracting many researchers and engineers for the last three decades.
The main reason is that the nonholonomic systems do not satisfy Brockett's theorem~\cite{brockett}.
The challenging and negative fact means that there is not any smooth time-invariant feedback control law to be able to stabilize them.
The applications include various types of robotic vehicles and manipulation.
Some of them have been often used as a kind of benchmark platform to demonstrate the performance of a proposed controller for not only a control problem of a single robotic system and also a distributed control problem of multiagent robotic systems.

The class subject to acceleration constraints---called second-order nonholonomic systems---includes real examples such as a V/STOL aircraft~\cite{vtol}, an underactuated manipulator~\cite{manipulator}, an underactuated hovercraft~\cite{hovercraft}, and a crane~\cite{crane}.
These systems can be represented in a canonical system called the second-order chained form by coordinate and input transformations.
The second-order chained form system is also affected by Brockett's theorem~\cite{brockett}.
To avoid this difficulty, there are several ingenious control approaches.
The stabilizing controllers proposed in  \cite{ge,hovercraft,pettersen_vessel,pettersen_vehicle} exploit discontinuity or time-variance; \cite{manipulator}, \cite{deluca} and \cite{aneke_ijrnc03} reduce the control problem into a trajectory tracking problem.
Other than those, \cite{yoshikawa} and \cite{ito_electron19} consider a motion planning problem (in other words, a feedforward control problem).

For the second-order chained form system, this paper presents a novel control approach composed of sinusoidal reference trajectories and a simple trajectory tracking controller.
The second-order chained form system is decomposed into three subsystems.
Two of them are the so-called double integrators; the other subsystem is a nonlinear system depending on one of the double integrators.
The double integrator is linearly controllable, which enables to transit the value of the position state in order to modify the nature of the nonlinear subsystem.
Transiting the value into ``one'' corresponds to modifying the nonlinear subsystem into the double integrator; transiting the value into ``zero'' corresponds to modifying the nonlinear subsystem into a linear autonomous system.
Focusing on this nature, 
this paper proposes a feedforward control strategy.
Furthermore, from the perspective of practical usefulness, the control strategy is extended into trajectory tracking control by using proportional-derivative (PD) feedback.

The remainder of this paper is organized as follows:
Section~\ref{sec:socf+syst-decomp-via-st-trans} presents that the second-order chained form system can be decomposed to linear subsystems by using state transitions.
On the basis of such system nature, Section~\ref{sec:proposal} proposes a feedforward control strategy and also a trajectory tracking controller of PD feedback.
Section~\ref{sec:experiment} applies the proposed control approach to an underactuated manipulator and evaluates it through numerical experiments.
The last section concludes the paper with a summary and future work.

\section{Subsystem Decomposition of The Second-Order Chained Form System By Using State Transitions}
\label{sec:socf+syst-decomp-via-st-trans}

Consider the following second-order chained form system:
\begin{equation}
  \label{eq:2cf}
  \frac{\diff^2}{\diff t^2}\,\bm{\xi}
  =
  \left[\begin{array}{@{\,}cc@{\,}}
  1 & 0\\
  0 & 1\\
  \xi_2 & 0
  \end{array}\right] \bm{u},
\end{equation}
where $\bm{\xi}=[\xi_1, \xi_2, \xi_3]^\top$ and $\bm{u}=[u_1, u_2]^\top$ are the generalized coordinate vector and the generalized input vector, respectively.
This system is well-known as a canonical form for a class of second-order nonholonomic systems, which can be resulted from the original  dynamical model via an appropriate transformation of the generalized coordinates and control inputs.
Representing the system~\eqref{eq:2cf} as an affine nonlinear system:
\begin{equation}
  \label{eq:affine}
  \frac{\diff}{\diff t}
  \left[\begin{array}{@{\,}c@{\,}}
  \xi_1\\ \xi_2\\ \xi_3\\ \dot{\xi}_1\\ \dot{\xi}_2\\ \dot{\xi}_3 
  \end{array}\right]
  =
  \left[\begin{array}{@{\,}c@{\,}}
  \dot{\xi}_1\\ \dot{\xi}_2\\ \dot{\xi}_3\\ 0\\ 0\\ 0
  \end{array}\right]
  +
  \left[\begin{array}{@{\,}c@{\,}}
  0\\ 0\\ 0\\ 1\\ 0\\ \xi_2
  \end{array}\right] u_1
  +
  \left[\begin{array}{@{\,}c@{\,}}
  0\\ 0\\ 0\\ 0\\ 1\\ 0
  \end{array}\right] u_2,
\end{equation}
we can easily confirm that the equilibrium points $(\xi_1^\star, \xi_2^\star, \xi_3^\star, 0, 0, 0),\ \xi_1^\star, \xi_2^\star, \xi_3^\star \in \mathbb{R}$ are small-time local controllable (STLC) via Sussmann's theorem~\cite{sussmann}.

By focusing on the control inputs, the system~\eqref{eq:2cf} can be decomposed into the following two subsystems:
\begin{subequations}
    \begin{align}
    \label{eq:u1}
    \frac{\diff}{\diff t}
    \left[\begin{array}{@{\,}c@{\,}}
      \xi_{1}\\ \xi_{3}\\ \dot{\xi}_{1}\\ \dot{\xi}_{3}
    \end{array}\right]
    &=
    \left[\begin{array}{@{\,}cccc@{\,}}
    0 & 0 & 1 & 0\\
    0 & 0 & 0 & 1\\
    0 & 0 & 0 & 0\\
    0 & 0 & 0 & 0
    \end{array}\right]
    \left[\begin{array}{@{\,}c@{\,}}
    \xi_{1}\\ \xi_{3}\\ \dot{\xi}_{1}\\ \dot{\xi}_{3}
    \end{array}\right]
    +
    \left[\begin{array}{@{\,}c@{\,}}
    0\\ 0\\ 1\\ \xi_{2}
    \end{array}\right] u_{1},
    \\
    \label{eq:u2}
    \frac{\diff}{\diff t} \left[\begin{array}{@{\,}c@{\,}}
    \xi_{2}\\ \dot{\xi}_{2}
    \end{array}\right]
    &=
    \left[\begin{array}{@{\,}cc@{\,}}
    0 & 1\\ 0 & 0
    \end{array}\right]
    \left[\begin{array}{@{\,}c@{\,}}
    \xi_{2}\\ \dot{\xi}_{2}
    \end{array}\right]
    +
    \left[\begin{array}{@{\,}c@{\,}}
    0 \\1
    \end{array}\right] u_{2}.
    \end{align}
\end{subequations}
The subsystem~\eqref{eq:u2} with respect to the control input~$u_2$ is a linear and controllable system represented by the double integrator.
On the other hand, the subsystem~\eqref{eq:u1} with respect to the input $u_1$ is a four-dimensional nonlinear system whose input matrix depends on the state variable $\xi_2$.
The subsystem~\eqref{eq:u1} can be further decomposed as follows:
\begin{subequations}
    \begin{align}
    \label{eq:sigma1}
    \frac{\diff}{\diff t}
    \left[\begin{array}{@{\,}c@{\,}}
    \xi_{1}\\ \dot{\xi}_{1}
    \end{array}\right]
    &=
    \left[\begin{array}{@{\,}cc@{\,}}
    0 & 1\\
    0 & 0
    \end{array}\right]
    \left[\begin{array}{@{\,}c@{\,}}
    \xi_{1}\\ \dot{\xi}_{1}
    \end{array}\right]
    +
    \left[\begin{array}{@{\,}c@{\,}}
    0\\ 1
    \end{array}\right] u_{1},
    \\
    \label{eq:sigma3}
    \frac{\diff}{\diff t}\left[\begin{array}{c}
    \xi_{3}\\ \dot{\xi}_{3}
    \end{array}\right]
    &=
    \left[\begin{array}{@{\,}cc@{\,}}
    0 & 1\\ 0 & 0
    \end{array}\right]
    \left[\begin{array}{@{\,}c@{\,}}
    \xi_{3}\\ \dot{\xi}_{3}
    \end{array}\right]
    +
    \left[\begin{array}{@{\,}c@{\,}}
    0 \\ \xi_{2}
    \end{array}\right] u_{1}.
    \end{align}
\end{subequations}
The subsystem~\eqref{eq:sigma1} of the double integrator is linear and controllable; the subsystem~\eqref{eq:sigma3} inherits the nonlinearity of the system~\eqref{eq:u1}.

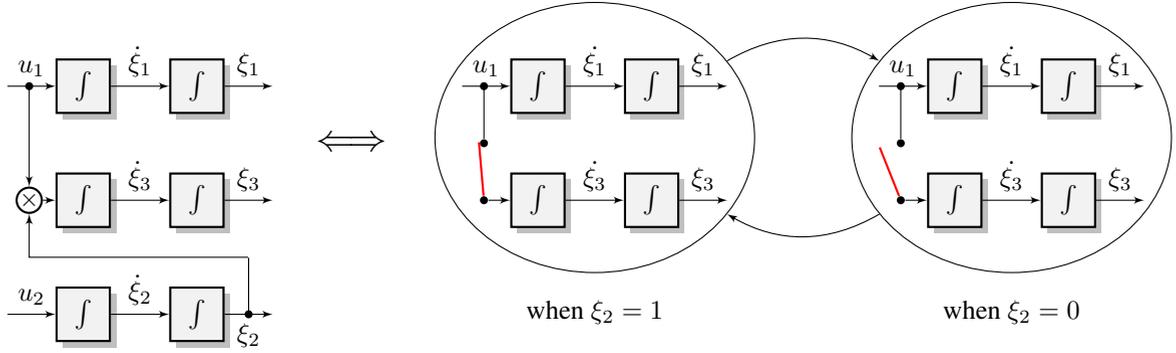
\begin{figure*}[t]
  \centering
  \begin{tikzpicture}[node distance=1cm,auto,>=latex']
    \node [int] (ss1int1) {$\int$};
    \node (u1) [left of=ss1int1, coordinate] {ss1int1};
    \node [right of=ss1int1, node distance=1.5cm] (ss1int2) [int] {$\int$};
    \node [coordinate] (z1) [right of=ss1int2]{};
    \path[->] (u1) edge node {$u_1$} (ss1int1);
    \path (u1) -- coordinate[xshift=-.03cm] (u1branch) (ss1int1);
    \path[->] (ss1int1) edge node {$\dot{\xi}_1$} (ss1int2);
    \draw[->] (ss1int2) edge node {$\xi_1$} (z1);
    \node[multp] (p1) [below of = u1branch, node distance = 1.5cm] {\small $\times$};
    \node [int] (ss2int1) [below of=ss1int1,node distance=1.5cm] {$\int$};
    \node [int] (ss2int2) [right of=ss2int1, node distance=1.5cm] {$\int$};
    \node [coordinate] (z3) [right of=ss2int2]{};
    \draw[->] (u1branch) node[branch] {}{} -- (p1);
    \draw[->] (p1) edge node {} (ss2int1);
    \path[->] (ss2int1) edge node {$\dot{\xi}_3$} (ss2int2);
    \draw[->] (ss2int2) edge node {$\xi_3$} (z3);
    \node [int] (ss3int1) [below of=ss2int1,node distance=1.5cm] {$\int$};
    \node (u2) [left of=ss3int1,node distance=1cm, coordinate] {ss3int1};
    \node [int] (ss3int2) [right of=ss3int1, node distance=1.5cm] {$\int$};
    \node [coordinate] (z2) [right of=ss3int2]{};
    \path (ss3int2) -- coordinate (z2branch) (z2);
    \path[->] (u2) edge node {$u_2$} (ss3int1);
    \path[->] (ss3int1) edge node {$\dot{\xi}_2$} (ss3int2);
    \draw[->] (ss3int2) edge node[below] {$\xi_2$} (z2);
    \node [branch, minimum size=0pt] (p3) [above of = z2branch, node distance = .75cm] {};
    \draw[->] (z2branch) node[branch] {}{} -- (p3) -| (p1);
      \node [int] (ss1int1-1) [right of = ss1int2, node distance = 4.5cm] {$\int$};
      \node (u1-1) [left of=ss1int1-1, coordinate] {ss1int1-1};
      \node [right of=ss1int1-1, node distance=1.5cm] (ss1int2-1) [int] {$\int$};
      \node [coordinate] (z1-1) [right of=ss1int2-1]{};
      \path[->] (u1-1) edge node {$u_1$} (ss1int1-1);
      \path (u1-1) -- coordinate[xshift=-.03cm] (u1branch-1) (ss1int1-1);
      \path[->] (ss1int1-1) edge node {$\dot{\xi_1}$} (ss1int2-1);
      \draw[->] (ss1int2-1) edge node {$\xi_1$} (z1-1);
      \node[branch] (p1-1) [below of = u1branch-1, node distance = .75cm] {};
      \draw (u1branch-1) node[branch] {}{} -- (p1-1);
      \node[above left=-0.15cm and .01cm of ss1int1-1.north west] (dummy_ss1-1) {};
      \node [int] (ss2int1-1) [below of=ss1int1-1,node distance=1.5cm] {$\int$};
      \node [int] (ss2int2-1) [right of=ss2int1-1, node distance=1.5cm] {$\int$};
      \node [coordinate] (z3-1) [right of=ss2int2-1, node distance=1cm]{};
      \path[->] (ss2int1-1) edge node {$\dot{\xi_3}$} (ss2int2-1);
      \draw[->] (ss2int2-1) edge node {$\xi_3$} (z3-1);
      \node[branch] (p2-1) [below of = u1branch-1, node distance = 1.5cm] {};
      \draw[->] (p2-1) -- (ss2int1-1);
      \node[branch,opacity=0] (dummy) [above left of = p1-1, node distance = .1cm] {};
      \path[thick,color=red] (p2-1) edge node {} (dummy);
      \node[below right=-0.3cm and .01cm of ss2int2-1.south east] (dummy_ss2-1) {};
      \node[draw,ellipse,fit={(dummy_ss1-1)(dummy_ss2-1)}] (plate_1) {};
      \node [int] (ss1int1-0) [right of = ss1int2-1, node distance = 4cm] {$\int$};
      \node (u1-0) [left of=ss1int1-0, coordinate] {ss1int1-0};
      \node [right of=ss1int1-0, node distance=1.5cm] (ss1int2-0) [int] {$\int$};
      \node [coordinate] (z1-0) [right of=ss1int2-0]{};
      \path[->] (u1-0) edge node {$u_1$} (ss1int1-0);
      \path (u1-0) -- coordinate[xshift=-.03cm] (u1branch-0) (ss1int1-0);
      \path[->] (ss1int1-0) edge node {$\dot{\xi_1}$} (ss1int2-0);
      \draw[->] (ss1int2-0) edge node {$\xi_1$} (z1-0);
      \node[branch] (p1-0) [below of = u1branch-0, node distance = .75cm] {};
      \draw (u1branch-0) node[branch] {}{} -- (p1-0);
      \node[above left=-0.15cm and .01cm of ss1int1-0.north west] (dummy_ss1-0) {};
      \node [int] (ss2int1-0) [below of=ss1int1-0,node distance=1.5cm] {$\int$};
      \node [int] (ss2int2-0) [right of=ss2int1-0, node distance=1.5cm] {$\int$};
      \node [coordinate] (z3-0) [right of=ss2int2-0, node distance=1cm]{};
      \path[->] (ss2int1-0) edge node {$\dot{\xi_3}$} (ss2int2-0);
      \draw[->] (ss2int2-0) edge node {$\xi_3$} (z3-0);
      \node[branch] (p2-0) [below of = u1branch-0, node distance = 1.5cm] {};
      \draw[->] (p2-0) -- (ss2int1-0);
      \node[branch,opacity=0] (dummy) [left of = p1-0, node distance = .3cm] {};
      \path[thick,color=red] (p2-0) edge node {} (dummy);
      \node[below right=-0.3cm and .01cm of ss2int2-0.south east] (dummy_ss2-0) {};
      \node[draw,ellipse,fit={(dummy_ss1-0)(dummy_ss2-0)}] (plate_0) {};
      \path[->] (plate_1) edge [bend left] node {} (plate_0);
      \path[->] (plate_0) edge [bend left] node {} (plate_1);
    \node [left of = p1-1, node distance=1.75cm] {{\Large $\Longleftrightarrow$}};
    \node [below of = plate_1, node distance=2.3cm] {when $\xi_2 = 1$};
    \node [below of = plate_0, node distance=2.3cm] {when $\xi_2 = 0$};
  \end{tikzpicture}
  \caption{Subsystem decomposition of the second-order chained form by using $\xi_2$'s state transitions between $0$ and $1$.}
  \label{fig:switch-image}
\end{figure*}
Fig.~\ref{fig:switch-image} shows a block diagram describing the above-mentioned subsystem decomposition explicitly.
The state of the subsystem~\eqref{eq:u2} can be transited to be a constant value because of the linear controllability.
For example, by setting time intervals where $\xi_2$ is ``zero'' and also $\xi_2$ is ``one'', the nonlinear subsystem~\eqref{eq:sigma3} can be treated as a linear system. During the time interval of $\xi_2 = 1$, the subsystems~\eqref{eq:sigma1} and~\eqref{eq:sigma3} are linear which have the same double integrator structure and control input~$u_1$.
On the other hand, during the time interval of $\xi_2 = 0$, the subsystem~\eqref{eq:u1} becomes a linear autonomous (i.e., uncontrollable) system and the subsystem~\eqref{eq:sigma1} can be controlled independently from subsystem~\eqref{eq:sigma3} by the control input $u_1$.

\begin{rmk}
Some conventional approaches such as in \cite{nam_sice02}, \cite{aneke_ijrnc03} and \cite{hably_cdc14} exploit a  different subsystem decomposition that can decompose the system~\eqref{eq:2cf} as follows:
\begin{subequations}
    \begin{align}
    \label{eq:another-sigma1}
    \frac{\diff}{\diff t}\left[\begin{array}{c}
    \xi_1\\ \dot{\xi}_1
    \end{array}\right]
    &=
    \left[\begin{array}{@{\,}cc@{\,}}
    0 & 1\\ 0 & 0
    \end{array}\right]
    \left[\begin{array}{@{\,}c@{\,}}
    \xi_{1}\\ \dot{\xi}_{1}
    \end{array}\right]
    +
    \left[\begin{array}{@{\,}c@{\,}}
    0 \\ 1
    \end{array}\right] u_1,
    \\
    \label{eq:another-sigma2}
    \frac{\diff}{\diff t}
    \left[\begin{array}{@{\,}c@{\,}}
    \xi_2\\ \xi_3\\ \dot{\xi}_2\\ \dot{\xi}_3
    \end{array}\right]
    &=
    \left[\begin{array}{@{\,}cccc@{\,}}
    0 & 0 & 1 & 0\\
    0 & 0 & 0 & 1\\
    0 & 0 & 0 & 0\\
    u_1 & 0 & 0 & 0\\
    \end{array}\right]
    \left[\begin{array}{@{\,}c@{\,}}
    \xi_2\\ \xi_3\\ \dot{\xi}_2\\ \dot{\xi}_3
    \end{array}\right]
    +
    \left[\begin{array}{@{\,}c@{\,}}
    0\\ 0\\ 1\\ 0
    \end{array}\right] u_2.
    \end{align}
\end{subequations}
The subsystem~\eqref{eq:another-sigma1} is the same with \eqref{eq:sigma1}; the subsystem~\eqref{eq:another-sigma2} has a variable structure depending on $u_1$.
The subsystem~\eqref{eq:another-sigma2} is linear when $u_1$ is a non-zero constant, which reduces a control problem of the second-order chained form system into a simultaneous stabilizing problem of the two subsystems~\eqref{eq:another-sigma1} and \eqref{eq:another-sigma2}.
When $u_1$ becomes zero before the end of control, however, the subsystem~\eqref{eq:another-sigma2} will be uncontrollable with a pole at the origin and then the whole of the subsystem loses the controllability.
This subsystem decomposition, therefore, needs control in consideration with $u_1$.
\end{rmk}

\section{Proposed Control Approach}
\label{sec:proposal}

In this paper, a control task of a rest-to-rest motion is addressed.
For this task, the authors propose a control approach composed of sinusoidal reference trajectories and a trajectory tracking controller.
In particular, a feedforward control strategy that generates the reference trajectories exploits the system decomposition based on state transition described in the previous section.

The feedforward control strategy using system switching based on state transitions in $\xi_2$ is as follows:
\begin{enumerate}[label=\bfseries Step \arabic*,leftmargin=*,labelindent=1.5em,labelsep=1.0em]
    \item\label{enum:ffcs-step1}
      Transit $\xi_2$ from any initial value to 1 by using $u_1(t)=0,\ u_2(t)=q_2(t)$;
    \item\label{enum:ffcs-step2}
      Transit $\xi_3$ from any initial value to any desired value (in conjunction with it, $\xi_1$ is also driven) by using $u_1(t)=q_3(t),\ u_2(t)=0$;
    \item\label{enum:ffcs-step3}
      Transit $\xi_2$ from 1 to 0 by using $u_1(t)=0,\ u_2(t)=q_2(t)$;
    \item\label{enum:ffcs-step4}
      Transit $\xi_1$ from any value in \ref{enum:ffcs-step2} to any desired value by using $u_1(t)=q_1(t),\ u_2(t)=0$;
    \item\label{enum:ffcs-step5}
      Transit $\xi_2$ from 0 to any desired value by using $u_1(t)=0,\ u_2(t)=q_2(t)$.
\end{enumerate}
A control input in \textbf{Step}~$k \; (k = 1, 2, \dotsc, 5)$ is designed by an appropriate sinusoidal function~$q_i(t) \; (i = 1, 2, 3)$ without any feedback.
This control strategy is namely motion planning, which naturally cannot deal with disturbance.
Therefore, we provide a trajectory tracking controller that follow the reference trajectory.

Consider to drive the state variables $\xi_i(t), \dot{\xi}_i(t)$ of the system~\eqref{eq:2cf} by the following sinusoidal functions with period~$T=2\pi/\omega$ and amplitude~$a_k$:
\begin{equation}
    \label{eq:control-input}
    q_i(t) = a_k\omega^2\sin\omega t.
\end{equation}
Then, at time $t\,(\leq kT)$, trajectories of a subsystem with non-zero input are derived as
\begin{align}
    \label{eq:traj-xid}
    \dot{\xi}_i(t) &= \dot{\xi}_i((k-1)T) - a_k\omega\cos\omega t + a_k\omega,
    \\
    \label{eq:traj-xi}
    \xi_i(t) &= \xi_i((k-1)T) + \dot{\xi}_i((k-1)T) t\nonumber\\
    &\phantom{=}\; \mbox{}-\dot{\xi}_i((k-1)T) (k-1) T\nonumber\\
    &\phantom{=}\; \mbox{}- a_k\sin\omega t + a_k\omega t -  a_k (k-1)\omega T,
\end{align}
respectively, where $\xi_i((k-1)T)$ and $\dot{\xi}_i((k-1)T)$ are initial values of the state variables in \textbf{Step $k$}. Thus, at the end of $k$-th period ($t=kT$), the state transitions are represented as
\begin{align}
    \dot{\xi}_i(kT) &= \dot{\xi}_i((k-1)T),\\
    \xi_i(kT) &= \xi_i((k-1)T) + \dot{\xi}_i((k-1)T)T + 2\pi a_k,
\end{align}
which means that a displacement of $2\pi a_k$ on $\xi_i$ is obtained. This can be seen that the desired displacement is extracted by using the amplitude $a_k$ as a tuning parameter.

By setting the trajectories~\eqref{eq:control-input}, \eqref{eq:traj-xid}, \eqref{eq:traj-xi} as reference trajectories $q_i^\mathrm{ref}(t)$, $\xi_i^\mathrm{ref}(t)$, $\dot{\xi}_i^\mathrm{ref}(t)$, a PD feedback control system can be designed for trajectory tracking. A linear system of a double integrator can be represented in the following state-space form with the state~$\bm{z}_i=[\xi_i, \dot{\xi}_i]^\top$ and control input~$q_i$:
\begin{equation}
    \label{eq:linear-system}
    \dot{\bm{z}}_i
    =
    \underbrace{%
    \left[\begin{array}{@{\,}cc@{\,}}
    0 & 1\\
    0 & 0
    \end{array}\right]%
    }_{\bm{A}} \bm{z}_i
    +
    \underbrace{%
    \left[\begin{array}{@{\,}c@{\,}}
    0\\
    1
    \end{array}\right]%
    }_{\bm{b}} q_i(t, \bm{z}_i).
\end{equation}
In \textbf{Step $k$}, a feedback controller for trajectory tracking to $\bm{z}^\mathrm{ref}_i$ is given as follows:
\begin{equation}
    \label{eq:FB-input}
    q_i(t, \bm{z}_i) = q_i^\mathrm{ref}(t) + \bm{k} \, \bm{e}_i,
\end{equation}
where $\bm{e}_i \coloneqq \bm{z}^\mathrm{ref}_i - \bm{z}_i$ and $\bm{k} = [k_p, k_d]$ is a feedback gain matrix. The system~\eqref{eq:linear-system} yields the closed-loop system $\dot{\bm{e}}_i = (\bm{A} - \bm{b} \bm{k})\bm{e}_i$. By choosing the feedback gain $\bm{k}$ so that $(\bm{A} - \bm{b} \bm{k})$ is Hurwitz-stable, the closed-loop system is stabilized, that is, $\bm{z}_i$ tracks $\bm{z}^\mathrm{ref}_i$.

\section{Numerical Experiments}
\label{sec:experiment}

In this section, we evaluate the effectiveness of the proposed control approach through numerical experiments.

\begin{figure}[tb]
    \centering
    \includegraphics[width=.9\linewidth]{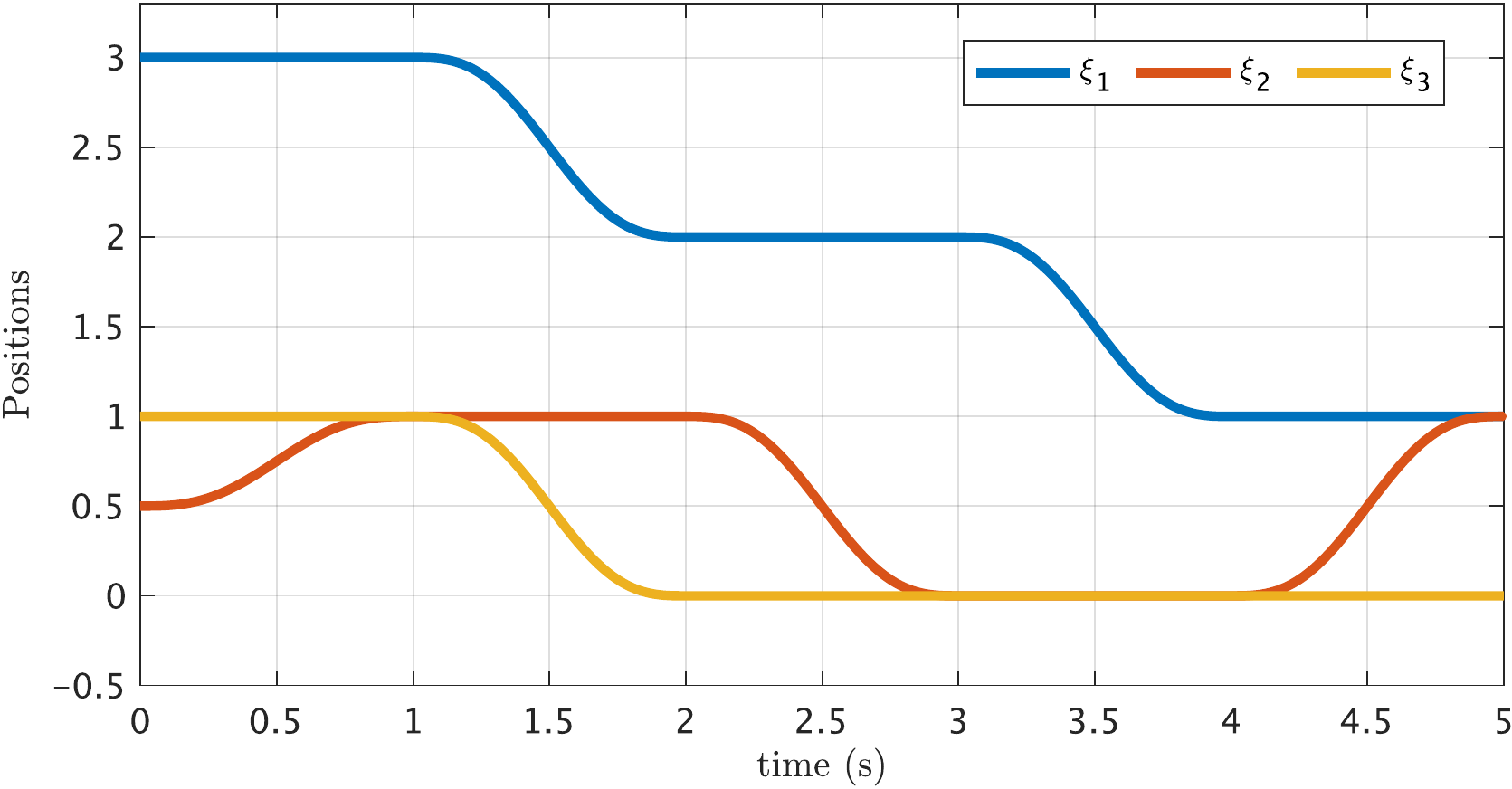}%
    \par\vspace{.75em}
    \includegraphics[width=.9\linewidth]{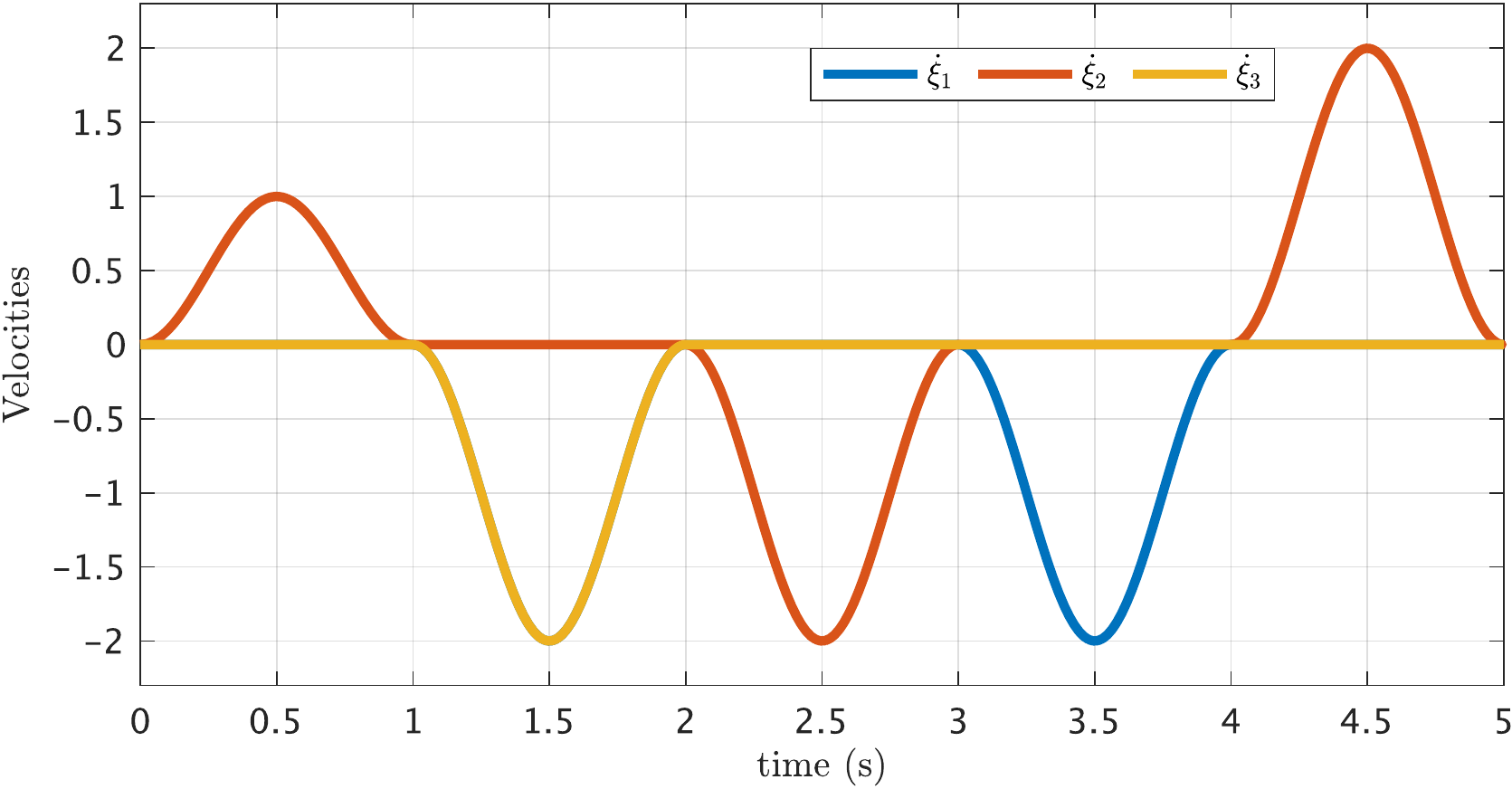}
    \par\vspace{.75em}\hspace{.2em}
    \includegraphics[width=.88\linewidth]{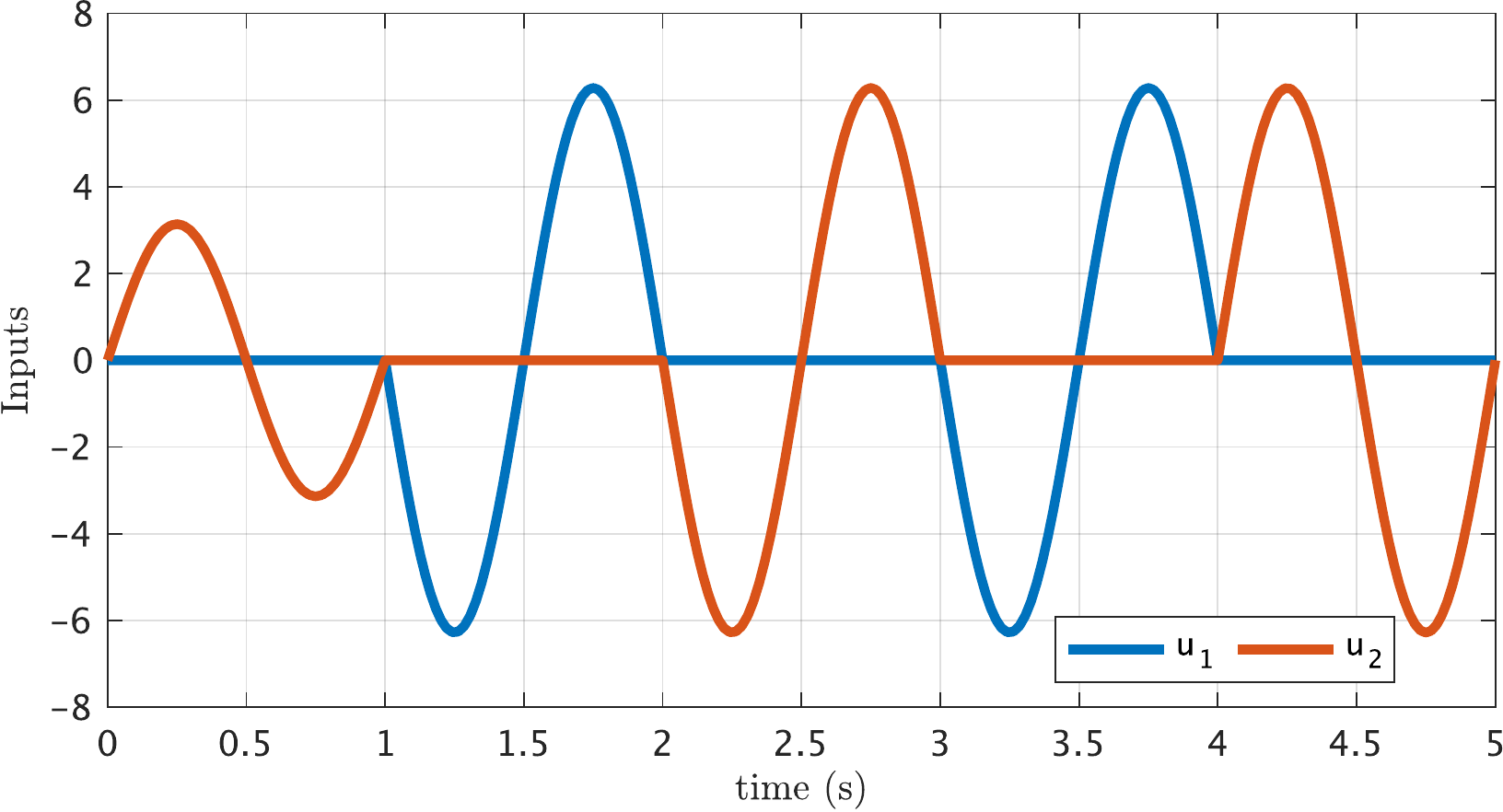}
    \caption{Simulation results of trajectory tracking control}
    \label{fig:FB}
\end{figure}
Firstly, we validate the proposed controller for the second-order chained form system.
A numerical experiment was performed with $T = 1\,\mbox{s}$, $\bm{\xi}(0) = [3, 0.5, 1]^\top$, $\dot{\bm{\xi}}(0) = \bm{0}_3$, $\bm{\xi}^\star=[1, 1, 0]^\top$, and $\dot{\bm{\xi}}^\star=\bm{0}_3$.
Fig.~\ref{fig:FB} shows the simulation results when choosing $a_1 = 1/(4\pi)$, $a_2 = a_3 = a_4 = -1/(2\pi)$, and $a_5 = 1/(2\pi)$. The ordinary differential equations was numerically solved by ODE45 of MATLAB~\cite{MATLAB_ODE45} with a relative tolerance of $1\times10^{-3}$.
The results indicate that each state reached to the target value $\bm{\xi}^\star$ with the remaining errors at $t = 5T$: $\bm{\xi}(5T)-\bm{\xi}^\star = [-2.7\times 10^{-8}, 1.0\times 10^{-10}, -4.7\times 10^{-8}]^\top$ and $\dot{\bm{\xi}}(5T)-\dot{\bm{\xi}}^\star = [1.3\times 10^{-8}, -8.9\times 10^{-9}, -2.1\times 10^{-8}]^\top$, which means that the desired control is achieved.

\begin{figure}[tb]
    \begin{center}
      \includegraphics[width=.9\linewidth]{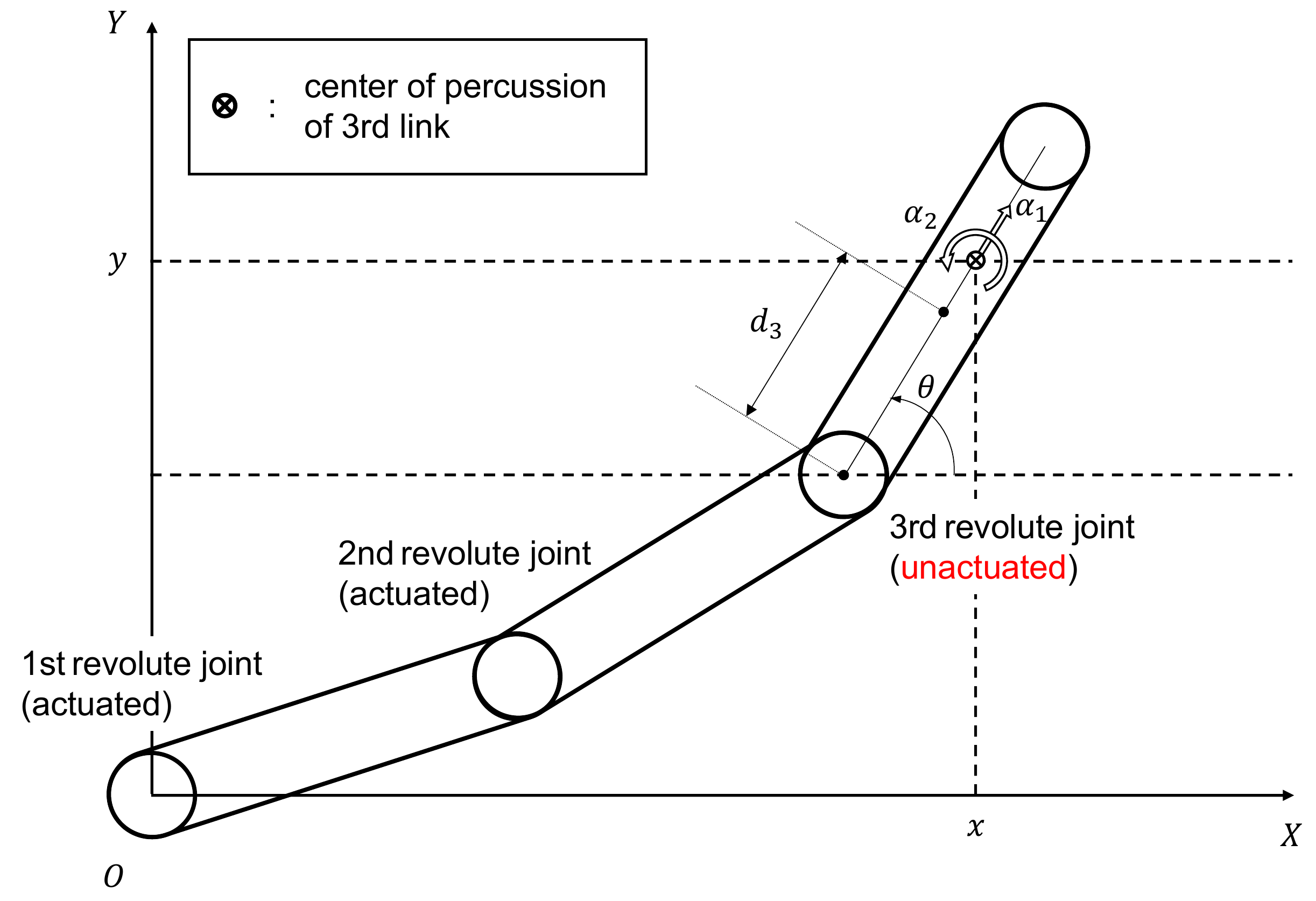}
    \end{center}
    \caption{A three-joint manipulator with passive third joint}
    \label{fig:manipulator}
\end{figure}
\begin{table*}[tb]
    \centering
    \caption{Definition of variables and parameters}
    \label{tab:manip-variable}
    \begin{tabular}{c@{\;}c@{\;\;}l}
      \hline
      & & \\[-.75em]
      $(x,y)$ & : & position of the center of percussion of the third link in the frame $O$-$XY$;
      \\[.25em]
      $\theta$ & : & angle of the third link relative to $X$-axis;
      \\[.25em]
      $d_3$ & : & distance between the third joint and the center of mass of the third link;
      \\[.25em]
      $m_3$ & : & mass of the third link;
      \\[.25em]
      $I_3$ & : & moment of inertia mass of the third link;
      \\[.25em]
      $L_{\mathrm{CoP}}$ & : & distance between the third joint and the center of percussion of the third link 
      $\left( L_{\mathrm{CoP}}\coloneqq (I_3+m_3d_3^2)/(m_3d_3)\right)$;
      \\[.25em]
      $\alpha_1$ & : & translational acceleration along the third link;
      \\[.25em]
      $\alpha_2$ & : & angular acceleration around the center of percussion of the third link.
      \\[.25em]
      \hline
    \end{tabular}
\end{table*}
Secondly, the proposed controller is applied to an underactuated manipulator---a typical example of second-order nonholonomic systems---as shown in Fig.~\ref{fig:manipulator}.
This manipulator has first two joints being actuated and the last joint being unactuated.
The system representation can be converted to the second-order chained form system.
Even if the third joint cannot be driven due to no actuator, the acceleration $(\alpha_1, \alpha_2)$ acting on the center of percussion of the third link can be treated equivalently as a control input owing to dynamic coupling effect---the rotational actuation of the first and second joints propagates through the links.
For simplicity, assume that there is no disturbance such as load, friction, linear and nonlinear damping, etc.
The main variables are defined as in Table~\ref{tab:manip-variable}.

Let $\bm{\chi} \coloneqq [x,y,\theta]^\top$ and $\bm{\alpha} = [\alpha_1,\,\alpha_2]^\top$.
Yoshikawa,~\textit{et al.}~\cite{yoshikawa} provided a set of coordinate and input transformations to convert the manipulator dynamics derived from the Lagrange's equation of motion into the following system representation:
\begin{equation}
    \ddot{\bm{\chi}}=\left[ \begin{array}{@{\,}cc@{\,}}
      \cos\theta & 0\\
      \sin\theta & 0\\
      0 & 1
    \end{array}\right]\bm{\alpha}.\label{eq:67}
\end{equation}
Using the coordinate transformation
\begin{equation}
    \label{eq:transX2Z}
    \left[ \begin{array}{@{\,}c@{\,}}
      \xi_1\\
      \xi_2\\
      \xi_3
    \end{array}\right] = \left[ \begin{array}{@{\,}c@{\,}}
      x-L_{\mathrm{CoP}}\\
      \tan\theta\\
      y
    \end{array}\right],\quad \left[ \begin{array}{@{\,}c@{\,}}
      \dot{\xi}_1\\
      \dot{\xi}_2\\
      \dot{\xi}_3
    \end{array}\right] = \left[ \begin{array}{@{\,}c@{\,}}
      \dot{x}\\
      \dot{\theta}\sec^2 \theta\\
      \dot{y}
    \end{array}\right]
\end{equation}
and the input transformation
\begin{equation}
    \left[ \begin{array}{@{\,}c@{\,}}
      \alpha_1\\
      \alpha_2
    \end{array}\right]
    = \left[\begin{array}{@{\,}c@{\,}}
      u_1\sec\theta\\
      u_2\cos^2\theta - 2\dot{\theta}^2\tan\theta
    \end{array}\right],
\end{equation}
the system~\eqref{eq:67} can be transformed into the second-order chained form system~\eqref{eq:2cf}. Note that both transformation are singular point at $\theta = \pm\pi/2$.

For the third joint of the underactuated manipulator with $m_3 = 0.6\,\mathrm{kg}$, $d_3 = 0.3\,\mathrm{m}$, and $I_3 = 4.5\times 10^{-3}\,\mathrm{kg\cdot m^2}$, steer from initial values $\bm{\chi}(0)=[3.33\,\mathrm{m}, 1\,\mathrm{m}, 4.6\times 10^{-1}\,\mathrm{rad}]^\top, \dot{\bm{\chi}}(0)=\bm{0}_3$ to the desired ones $\bm{\chi}^\star=[1\,\mathrm{m}, 0\,\mathrm{m}, 0\,\mathrm{rad}]^\top, \dot{\bm{\chi}}^\star=\bm{0}_3$.

\begin{figure*}[tb]
    \centering
    \begin{minipage}{.475\textwidth}
      \centering
      \includegraphics[width=.9\textwidth]{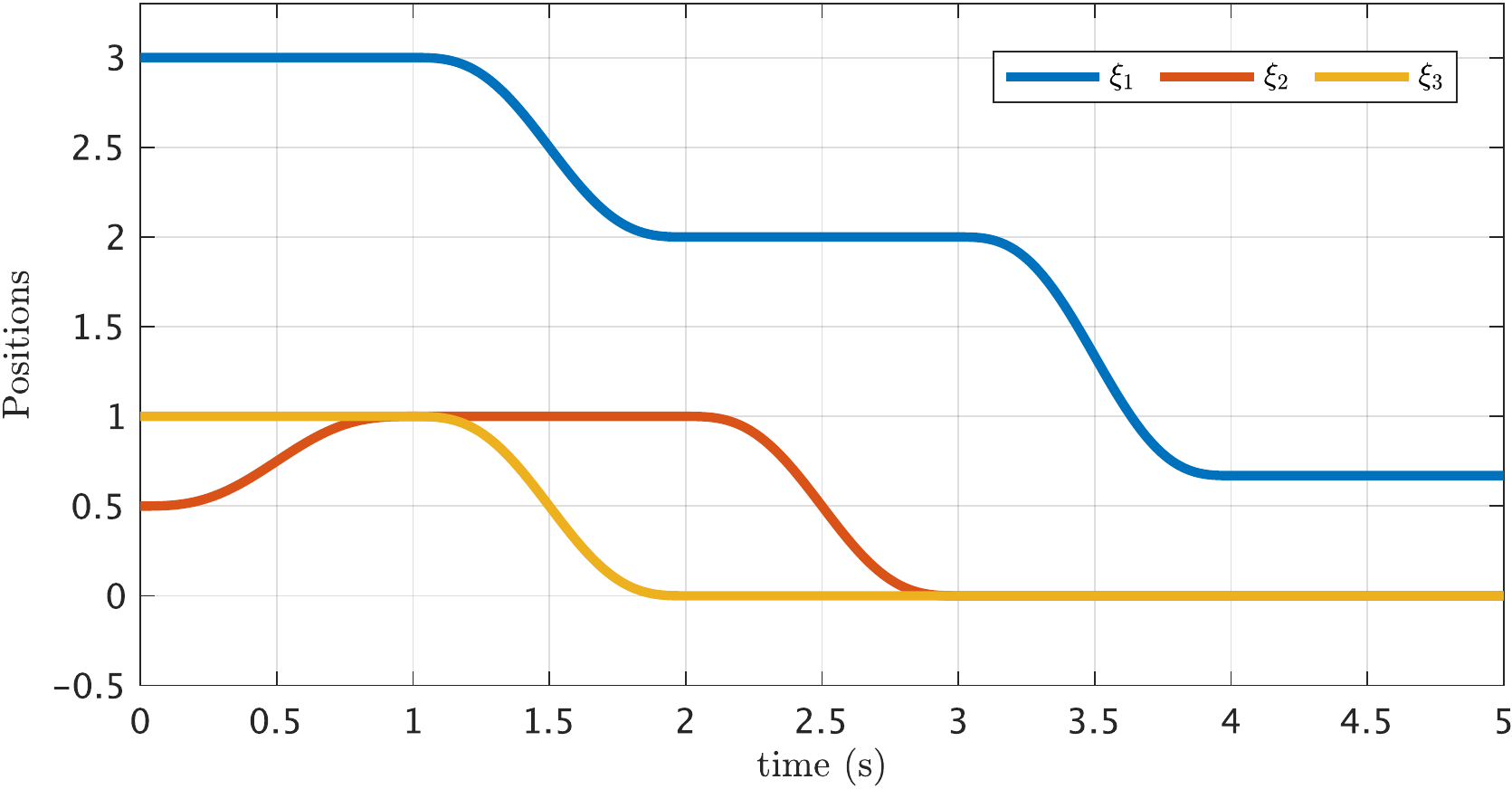}
      \par\vspace{.75em}
      \includegraphics[width=.9\textwidth]{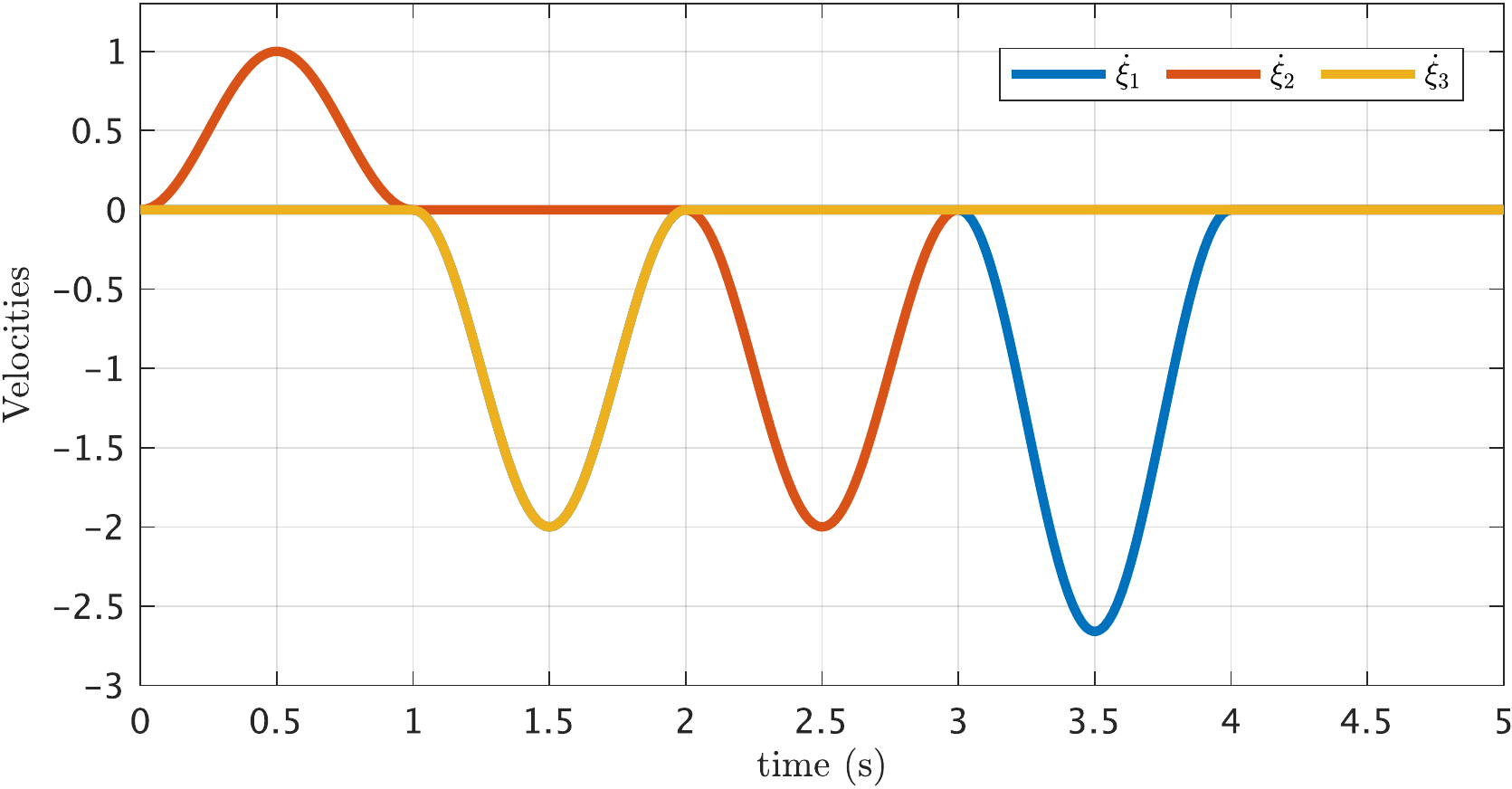}
      \par\vspace{.75em}
      \includegraphics[width=.9\textwidth]{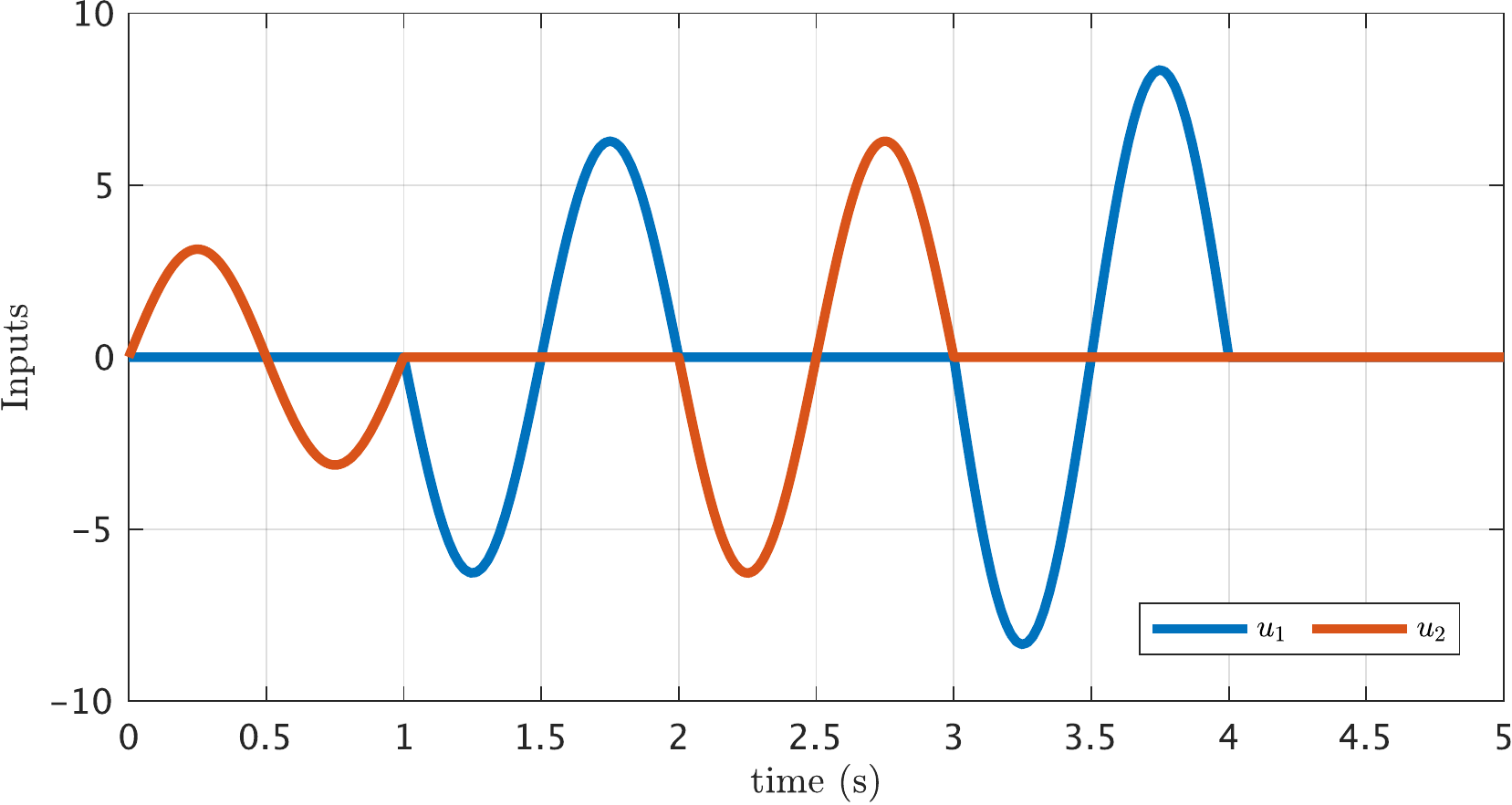}\par
      {\footnotesize (a) States and inputs of the second-order chained form system}
    \end{minipage}
    \begin{minipage}{.475\textwidth}
      \centering
      \includegraphics[width=.9\linewidth]{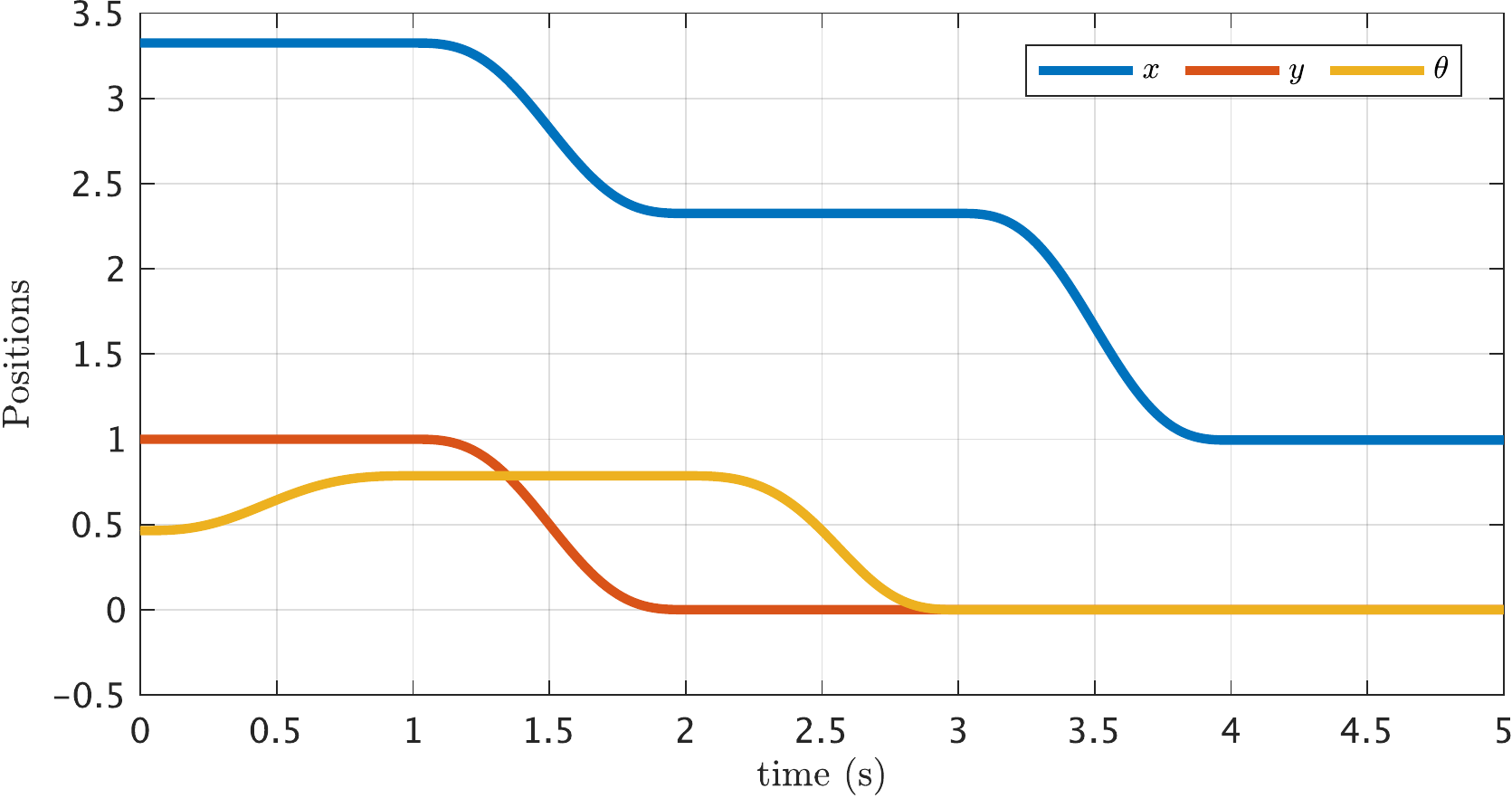}
      \par\vspace{.75em}
      \includegraphics[width=.9\linewidth]{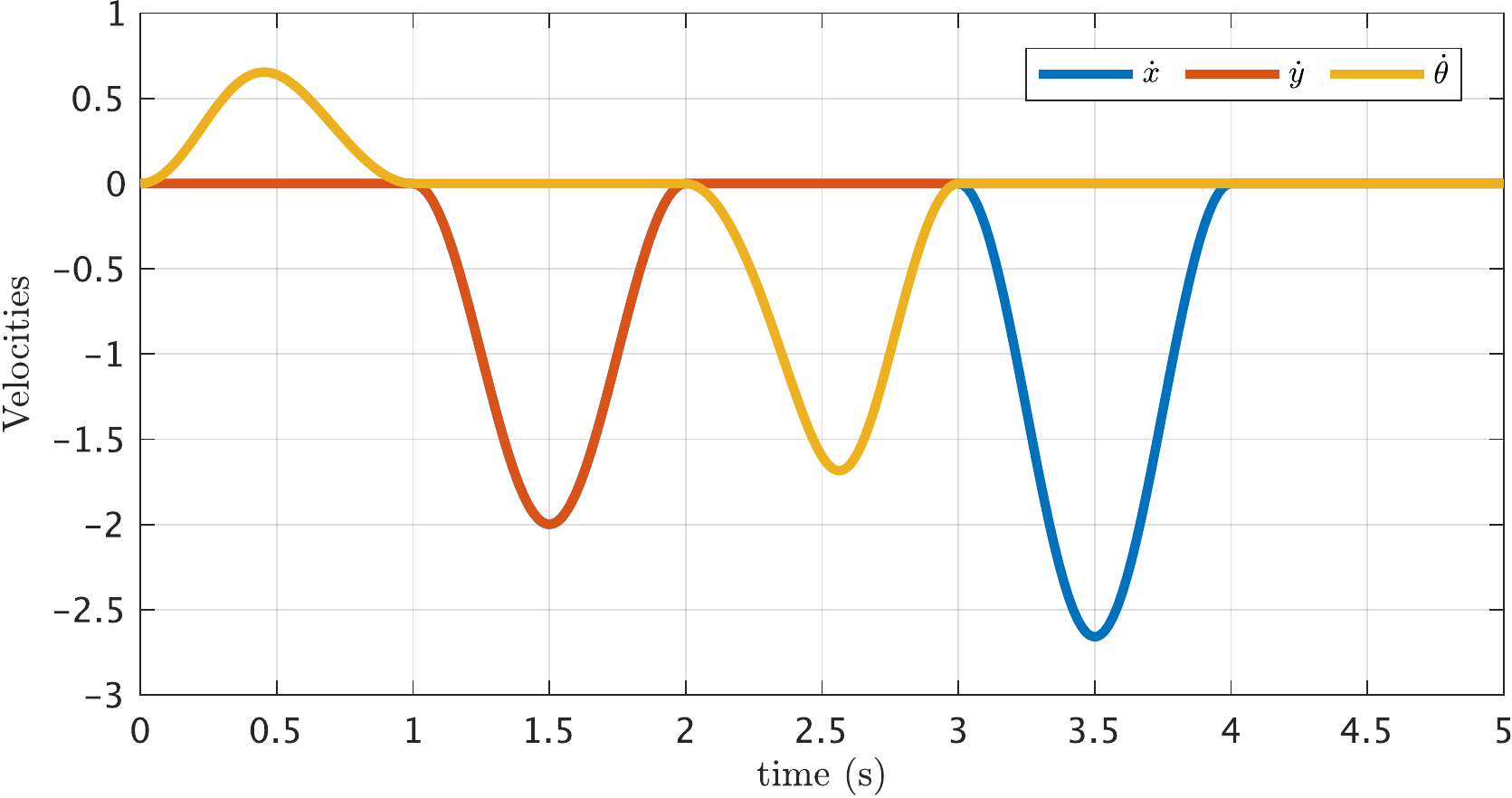}
      \par\vspace{.75em}
      \includegraphics[width=.9\linewidth]{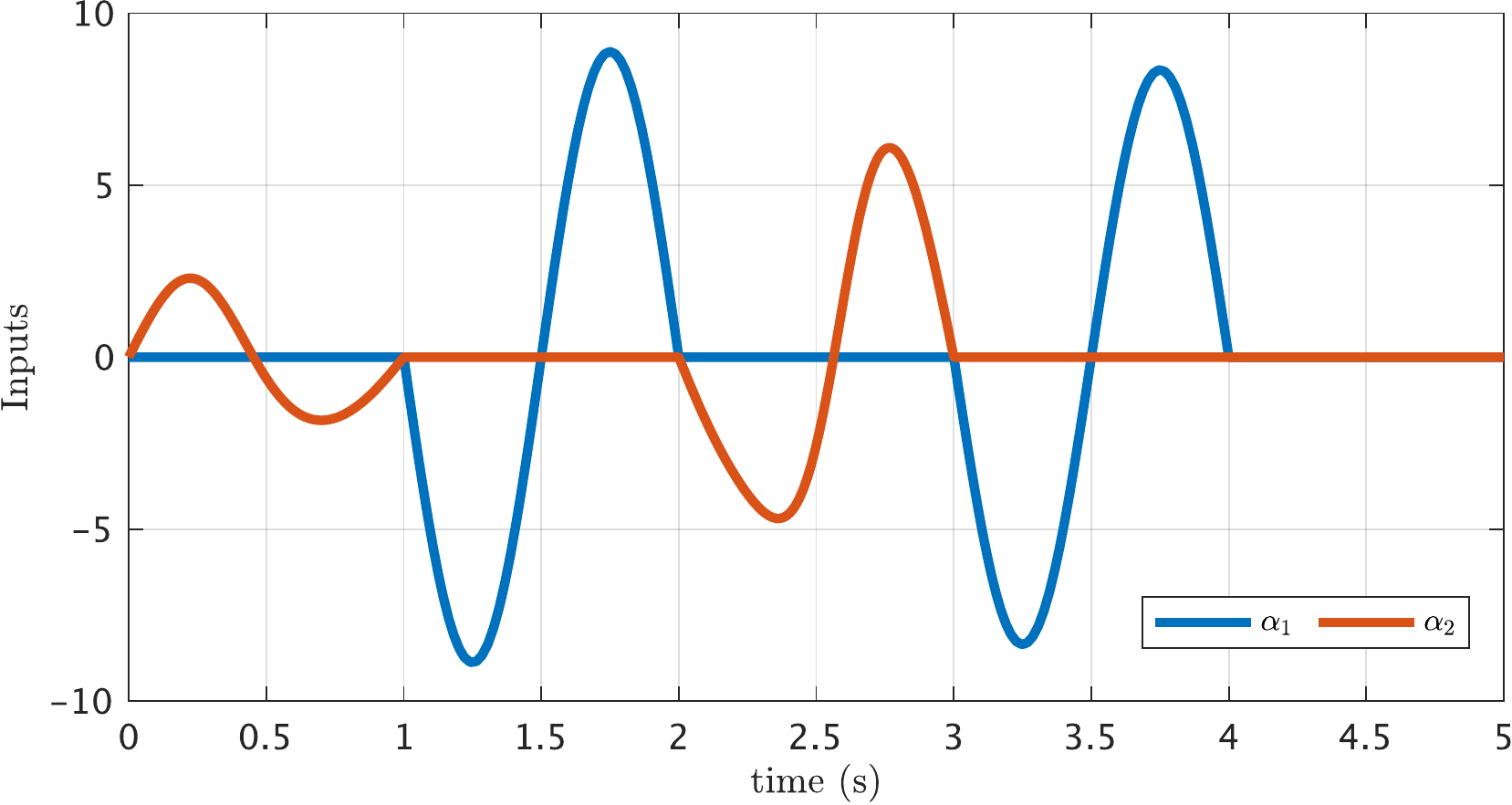}\par
      {\footnotesize (b) Status and inputs of a three-joint underactuated manipulator}
    \end{minipage}
    \caption{Numerical results}
    \label{fig:trans-manip}
\end{figure*}
Fig.~\ref{fig:trans-manip} shows a simulation result with the period $T=1\, \mathrm{s}$ and the feedback gain $k_p = k_d = 1$. In this case, from \eqref{eq:transX2Z}, we have $\bm{\xi}(0)=[3, 0.5, 1]^\top$ and $\bm{\xi}^\star=[0.67, 0, 0]^\top$. It can be confirmed that each state converges to the desired value in the both system representation.

\begin{figure}[tb]
    \centering
    \hspace{1.7em}
    \includegraphics[width=.876\linewidth]{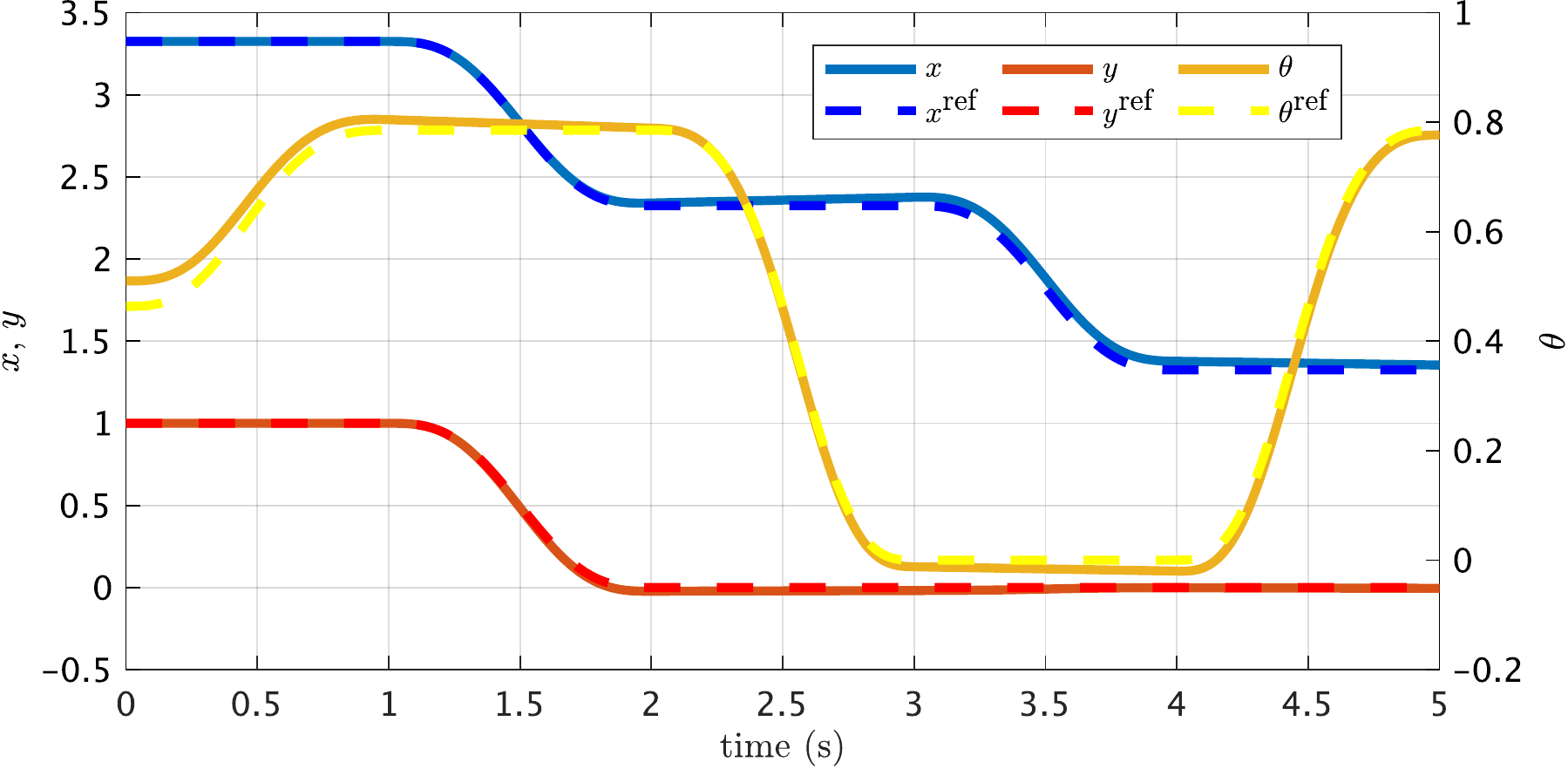}
    \par\vspace{.75em}\hspace{1.7em}
    \includegraphics[width=.877\linewidth]{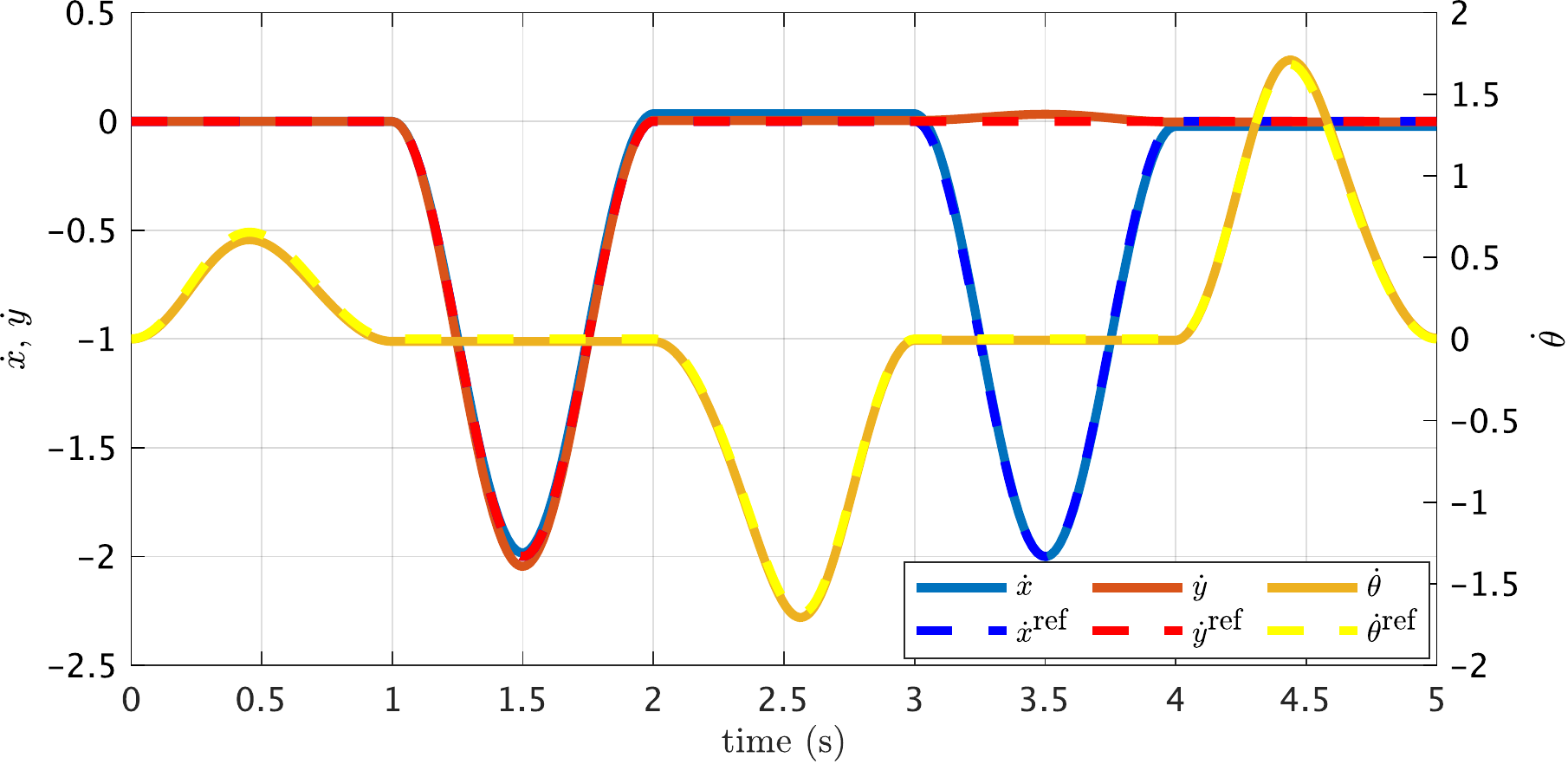}
    \par\vspace{.75em}\hspace{.2em}
    \includegraphics[width=.8\linewidth]{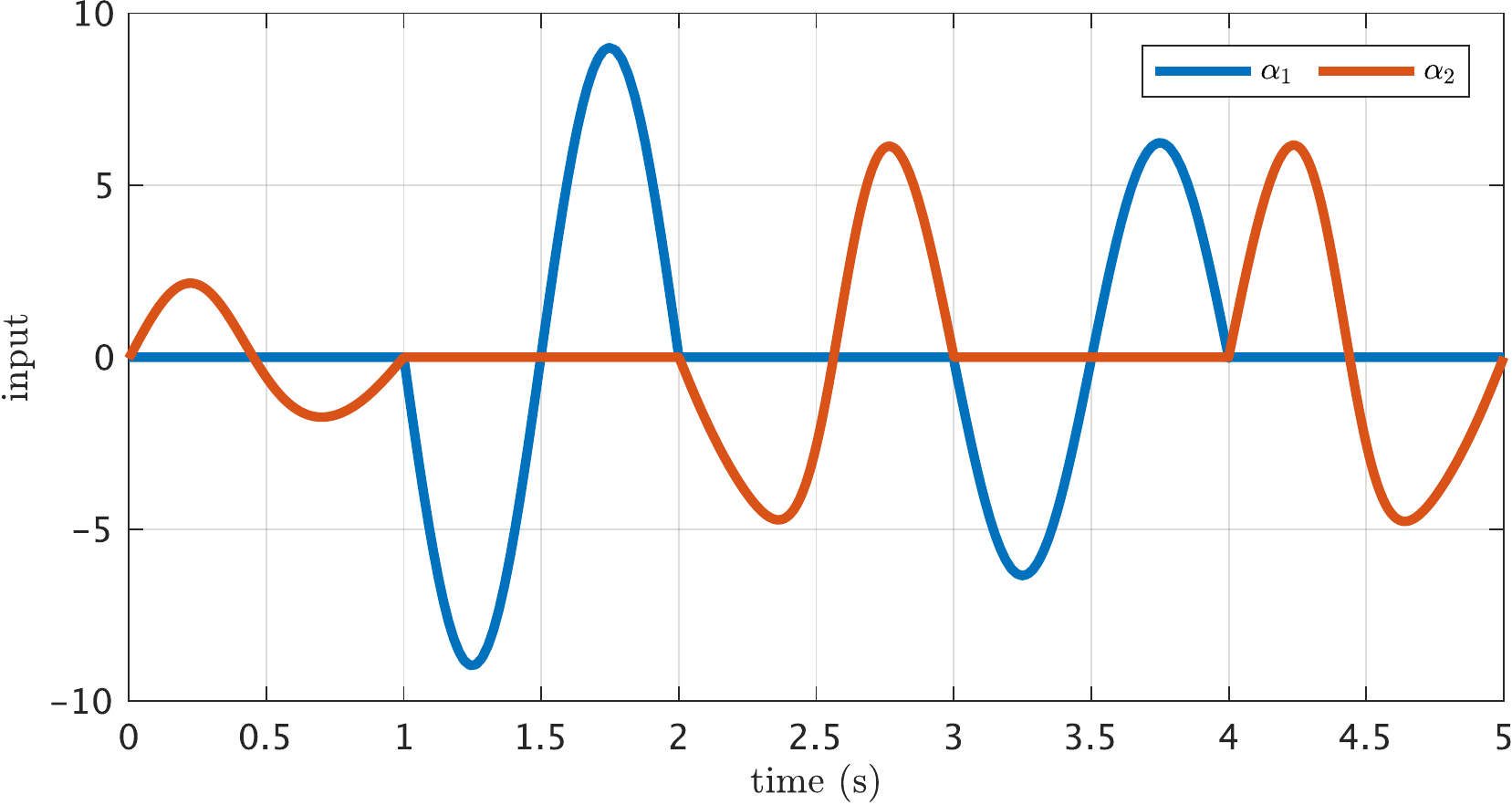}
    \par\vspace{.75em}\hspace{-.25em}
    \includegraphics[width=.82\linewidth]{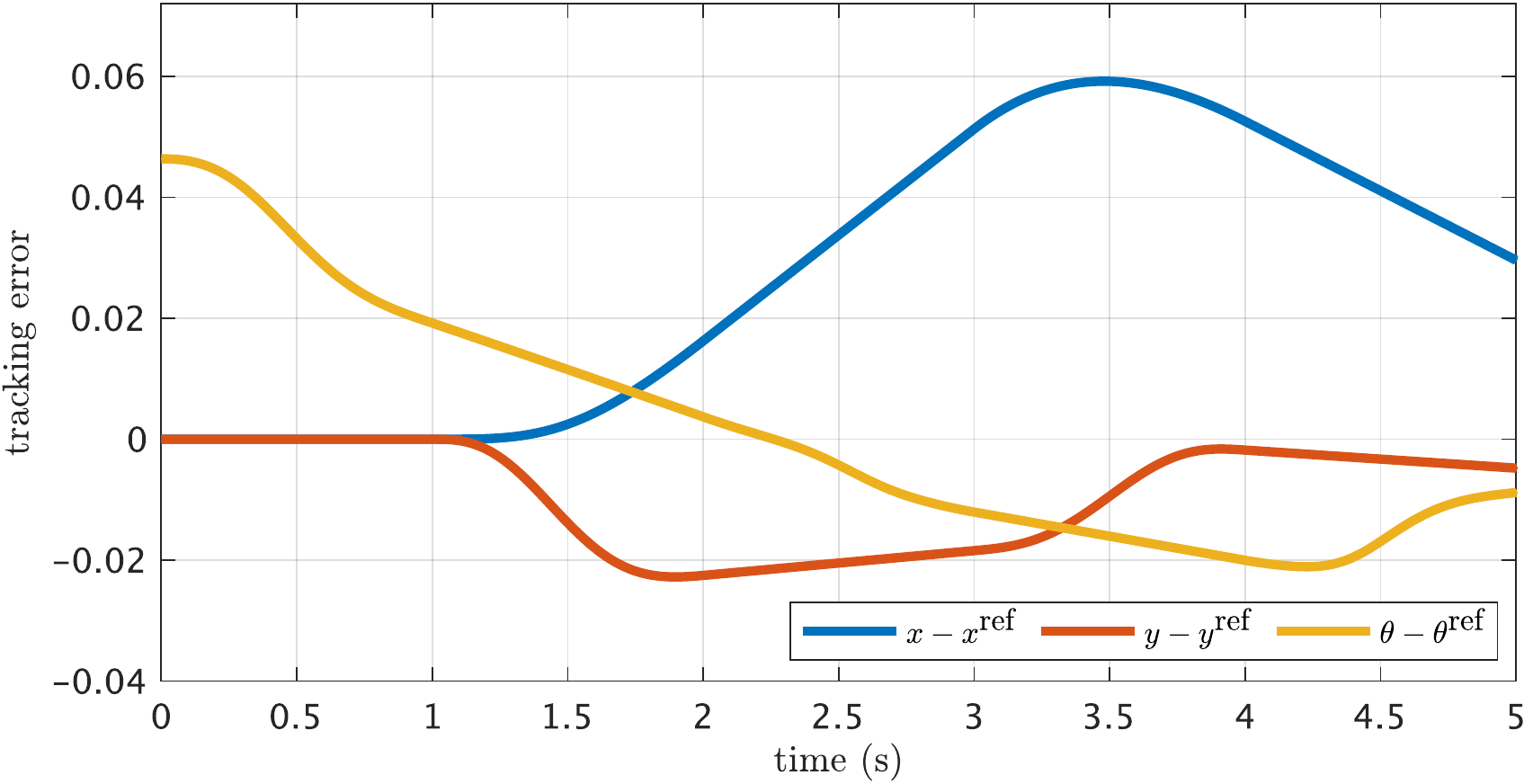}
    \caption{Given an initial value error$(+10\%)$}
    \label{fig:10per-error}
\end{figure}

Furthermore, to verify the effect of feedback control, another case with an initial value error was simulated. For a rest-to-rest motion from $\bm{\chi}(0)=[3.33\,\mathrm{m}, 1\,\mathrm{m}, 4.6\times 10^{-1}\,\mathrm{rad}]^\top$ to $\bm{\chi}^\star=[1.33\,\mathrm{m}, 0\,\mathrm{m}, 7.8\times 10^{-1}\,\mathrm{rad}]^\top$ with the zero velocities, the initial value error of $+10\%$ is given to $\theta$, i.e., $\bm{\chi}(0)=[3.33\,\mathrm{m}, 1\,\mathrm{m}, 5.1\times 10^{-1}\,\mathrm{rad}]^\top$. The result is shown in Fig.~\ref{fig:10per-error}. The dashed lines indicate the target trajectories. It can be observed that tracking error due to the initial value error is alleviated over time.

Similarly, when initial value errors of $\pm 1\%, \pm 10\%$, and $\pm 30\%$ on $\theta$ are given the tracking errors at the end of control at $t=5T$ are summarized in Table~\ref{tab:init-error}. The terminal values of the tracking errors do not increase greatly even if the magnitude of the initial value error increases. Consequently, it is confirmed that the feedback of trajectory tracking has a sufficient effect on initial value errors. Note that the terminal error on $x$ is relatively larger than the one on $\theta$.
The proposed control method attempts to settle the system by focusing on a single state every step.
In addition, the state in which the control step ends has no chance to be controlled directly.
For such a state, there can be a secondary state transition that yields in control steps that focus on the other states.
Therefore, if a state fails to converge into its reference trajectory within the control step due to initial value error or disturbance, it behaves unexpectedly until the end of the control strategy.
In particular, $\xi_2$---the state used for switching the systems---has a negative effect on the other states because the reference trajectory is not computed correctly.
Furthermore, the error remaining in the velocity state ($\xi_4, \xi_5, \xi_6$) causes a drift in the position state ($\xi_1, \xi_2, \xi_3$) even if the input is zero in the following control steps.
This is explained by numerical experiments shown in Fig.~\ref{fig:10per-error}.
Note that $\theta$ is related to $\xi_2$ as specified in \eqref{eq:transX2Z}.
This means that $\theta$ affects the other states $(x, y)$ when not converging completely.
On the other hand, since $\xi_2$ is settled in the final step (i.e., \ref{enum:ffcs-step5}), the propagation from the error in the velocity state is small.
Therefore, the error remaining in $\theta$ is considered to be smaller than in $x$.

\begin{table}[tb]
    \centering
    \caption{Error from target value by the initial value error}
    \label{tab:init-error}
    \begin{tabular}{@{\;}lc@{\:}c@{\;}}
        \multicolumn{1}{c}{Case} & $\bm{\chi}(0)-\bm{\chi}^\star$ & $\bm{\chi}(5T)-\bm{\chi}^\star$
        \\\hline
        & & \\[-1ex]
        w/o init.~err. & 
        $\left[ \begin{array}{@{\,}l@{\,}} 
          \phantom{-4.6\times 10\ \;}0\ \mathrm{m}\\
          \phantom{-4.6\times 10\ \;}0\ \mathrm{m}\\
          \phantom{-4.6\times 10\ \;}0\ \mathrm{rad}
        \end{array}\right]$ & 
        $\left[ \begin{array}{@{\,}l@{\,}}
          \phantom{-}4.5\times 10^{-8}\ \mathrm{m}\\
          \phantom{-}4.4\times 10^{-7}\ \mathrm{m}\\
          \phantom{-}4.9\times 10^{-9}\ \mathrm{rad} 
        \end{array}\right]$
        \\[5ex]
        w/ $+1\,\%$ init.~err. &
        $\left[ \begin{array}{@{\,}l@{\,}}
          \phantom{-4.6\times 10\ \;}0\ \mathrm{m}\\
          \phantom{-4.6\times 10\ \;}0\ \mathrm{m}\\
          \phantom{-}4.6\times 10^{-3}\ \mathrm{rad}
        \end{array}\right]$ & 
        $\left[ \begin{array}{@{\,}l@{\,}}
          \phantom{-}2.9\times 10^{-3}\ \mathrm{m}\\
          -6.9\times 10^{-4}\ \mathrm{m}\\
          -8.5\times 10^{-4}\ \mathrm{rad}
        \end{array}\right]$
        \\[5ex]
        w/ $-1\,\%$ init.~err. &
        $\left[ \begin{array}{@{\,}l@{\,}}
          \phantom{-4.6\times 10\ \;}\mbox{0}\ \mathrm{m}\\
          \phantom{-4.6\times 10\ \;}\mbox{0}\ \mathrm{m}\\
          -4.6\times 10^{-3}\ \mathrm{rad}
        \end{array}\right]$ &
        $\left[ \begin{array}{@{\,}l@{\,}}
          -2.9\times 10^{-3}\ \mathrm{m}\\
          \phantom{-}7.4\times 10^{-4}\ \mathrm{m}\\
          \phantom{-}8.5\times 10^{-4}\ \mathrm{rad}
        \end{array}\right]$
        \\[5ex]
        w/ $+10\,\%$ init.~err. &
        $\left[ \begin{array}{@{\,}l@{\,}}
          \phantom{-4.6\times 10\ \;}0\ \mathrm{m}\\
          \phantom{-4.6\times 10\ \;}0\ \mathrm{m}\\
          \phantom{-}4.6\times 10^{-2}\ \mathrm{rad}
        \end{array}\right]$ & 
        $\left[ \begin{array}{@{\,}l@{\,}}
          \phantom{-}3.0\times 10^{-2}\ \mathrm{m}\\
          -4.8\times 10^{-3}\ \mathrm{m}\\
          -8.8\times 10^{-3}\ \mathrm{rad}
        \end{array}\right]$
        \\[5ex]
        w/ $-10\,\%$ init.~err. & 
        $\left[ \begin{array}{@{\,}l@{\,}}
          \phantom{-4.6\times 10\ \;}0\ \mathrm{m}\\
          \phantom{-4.6\times 10\ \;}0\ \mathrm{m}\\
          -4.6\times 10^{-2}\ \mathrm{rad}
        \end{array}\right]$ &
        $\left[ \begin{array}{@{\,}l@{\,}}
          -2.9\times 10^{-2}\ \mathrm{m}\\
          \phantom{-}9.9\times 10^{-3}\ \mathrm{m}\\
          \phantom{-}8.3\times 10^{-3}\ \mathrm{rad}
        \end{array}\right]$
        \\[5ex]
        w/ $+30\,\%$ init.~err. &
        $\left[ \begin{array}{@{\,}l@{\,}}
          \phantom{-4.6\times 10\ \;}0\ \mathrm{m}\\
          \phantom{-4.6\times 10\ \;}0\ \mathrm{m}\\
          \phantom{-}1.4\times 10^{-1}\ \mathrm{rad}
        \end{array}\right]$ &
        $\left[ \begin{array}{@{\,}l@{\,}}
          \phantom{-}9.1\times 10^{-2}\ \mathrm{m}\\
          \phantom{-}3.3\times 10^{-3}\ \mathrm{m}\\
          -2.8\times 10^{-2}\ \mathrm{rad}
        \end{array}\right]$
        \\[5ex]
        w/ $-30\,\%$ init.~err. &
        $\left[ \begin{array}{@{\,}l@{\,}}
          \phantom{-4.6\times 10\ \;}0\ \mathrm{m}\\
          \phantom{-4.6\times 10\ \;}0\ \mathrm{m}\\
          -1.4\times 10^{-1}\ \mathrm{rad}
        \end{array}\right]$ &
        $\left[ \begin{array}{@{\,}l@{\,}}
          -8.5\times 10^{-2}\ \mathrm{m}\\
          \phantom{-}4.3\times 10^{-2}\ \mathrm{m}\\
          \phantom{-}2.3\times 10^{-2}\ \mathrm{rad}
        \end{array}\right]$
        \\[5ex]
        \hline
    \end{tabular}
\end{table}

\section{Conclusion}
\label{sec:conclusion}

In this paper, a novel control approach composed of sinusoidal reference trajectories and a simple trajectory tracking controller for the second-order chained form system was proposed.
The key idea is a subsystem decomposition of the second-order chained form system by using state transitions.
The effectiveness of the proposed algorithm was demonstrated by numerical results including an application to a three-joint underactuated manipulator.
In particular, it can be confirmed that the feedback control works well against the initial value error.

The future work of this research is to verify the proposed approach via experiments on an actual robot.


\begin{IEEEbiography}[{\includegraphics[width=1in,height=1.25in,clip,keepaspectratio]{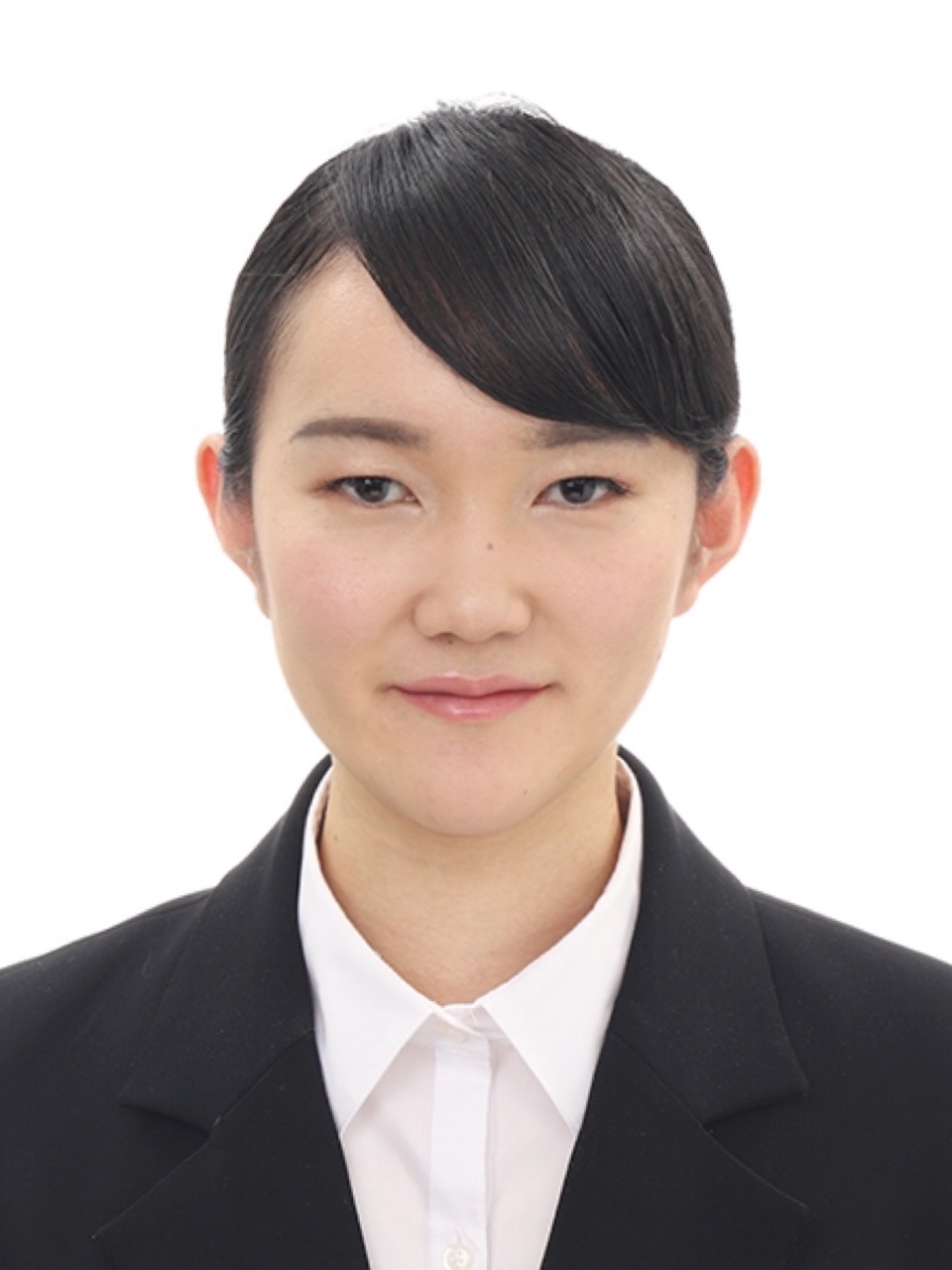}}]{Mayu Nakayama} was born in Kiyosu, Aichi, Japan in 
1997.  She received the B.S. and M.S. degrees in information science and technology from Aichi Prefectural University (APU), Nagakute, Aichi, Japan, in 2020 and 2022.

She is currently with DENSO Corporation.
Her research interests include nonlinear control for underactuated systems.
\end{IEEEbiography}

\begin{IEEEbiography}[{\includegraphics[width=1in,height=1.25in,clip,keepaspectratio]{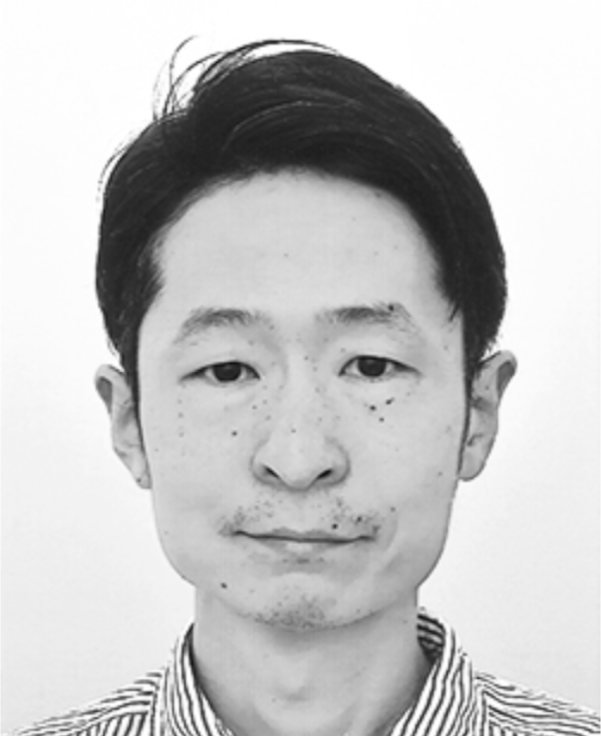}}]{Masahide Ito} (M'10) was born in Nagoya, Aichi, Japan in 
1979.  He received the B.S., M.S., and Ph.D. degrees in information science and technology from Aichi Prefectural University (APU), Nagakute, Aichi, Japan, in 2002, 2004, and 2008.

He is currently an Associate Professor with the School of Information Science and Technology, APU.
His research interests include visual feedback control of robotic systems and nonlinear control for underactuated systems.

\end{IEEEbiography}
\EOD

\end{document}